\documentclass[10pt,twocolumn,showpacs,pre,preprintnumbers,amsmath,amssymb,aps]{revtex4-1}
\usepackage{graphicx,latexsym,amsfonts,dcolumn,bm,color,enumerate}
\usepackage[percent]{overpic}
\newcommand{\snr}{S\hspace{-0.3ex}N\hspace{-0.3ex}R}
\newcommand{\us}[1]{_\textrm{\scriptsize #1}}
\newcommand{\Tn}{T\us{n}}
\newcommand{\Bn}{B\us{n}}
\newcommand{\Pn}{\mathbf{P}\us{n}}
\newcommand{\Cn}{C\us{n}}
\newcommand{\Ce}{C\us{e}}
\newcommand{\etal}{\textit{et al.}}
\begin{document}
%
%
\title{Swarming collapse under limited information flow between individuals}
\author{Mohammad Komareji$^{1}$}
\author{Yilun Shang$^{1,2}$} \email{shyl@tongji.edu.cn}
\author{Roland Bouffanais$^1$} \email{bouffanais@sutd.edu.sg}
\affiliation{$^1$Singapore University of Technology and Design, 8 Somapah
  Road, Singapore 487372\\$^2$ Department of Mathematics, Tongji University,
  Shanghai 200092, China}

\begin{abstract} The emergence of collective decision in swarms and their
  coordinated response to complex environments underscore the central role
  played by social transmission of information. Here, the different possible
  origins of information flow bottlenecks are identified. Using a combination
  of network-, control- and information-theoretic elements applied to a group
  of interacting self-propelled particles, the effect of varying information
  capacity of the signaling channel on dynamic collective behaviors is
  revealed. We find a sufficient condition on the information data rate that
  guarantees the effectiveness of swarming while also highlighting the
  profound connection with the topology of the underlying interaction
  network. We also show that when decreasing the data rate, the swarming
  behavior invariably vanishes following a second-order phase transition
  irrespective of the intrinsic noise level. The variations along the
  transition line are found to be in good agreement with information-theoretic
  predictions.
\end{abstract}
%
%
\maketitle

%

Information is a crucial currency for animals from a behavioral
standpoint. This has important consequences when considering collective
behaviors of interacting individuals~\cite{bouffanais}. Information exchange is critical to the
execution and effectiveness of a host of collective behaviors: fish schooling,
birds flocking, amoebae aggregating, locusts marching or more generally agents
swarming~\cite{%
  sumpter06:_princ_of_collec_animal_behav,%
  mccann10:_cell,hsieh08:_decen,bialek12:_statis,%
  attanasi14:_infor,bazazi08:_collec,shang14:_influen,%
  haque11:_biolog}.  However, animals are generally faced with uncertain
environments and fast-changing circumstances, and, often, their survival
critically depends on their ability to swiftly manage and respond to such
unpredictable changes in their surroundings. It is now widely believed that
the benefits of swarming are directly related to their enhanced ability in
dynamically responding to uncertain and rapidly-changing natural
environments~\cite{krause02:_livin_in_group,sumpter10:_collec_animal_behav}.
In recent years, there has been mounting recognition that distributed transfer
of behavioral information is key to the highly responsive nature of
swarms~\cite{strandburg-peshkin13:_visual,%
  sumpter08:_infor,lemasson13:_motion,%
  szabo09:_trans,bode10:_how,%
  handegard12:_dynam_coord_group_huntin_collec}.  It is now becoming apparent
that collective intelligence in the form of adaptive behavioral response
relies upon having both accurate and sufficient social information exchanges
occurring among interacting units.

Networked dynamical systems (NDSs) and multiagent adaptive systems are
engineering embodiments of natural swarms. A key problem with these systems is
the design of controls achieving specific collective behaviors in the presence
of limited or unreliable information exchanges and dynamically changing
topologies~\cite{hespanha07:_survey_recen_resul_networ_contr_system,%
  baillieul07:_contr_commun_chall_networ_real_time_system}. In the past two
decades, significant advances have been achieved paving the way to emerging
engineering applications such as the control of distributed sensor networks,
the coordination of autonomous---air, surface and underwater---vehicles,
robotic swarming,
etc.~\cite{olfati-saber07:_consen_cooper_networ_multi_agent_system}. Information
and communication constraints are now recognized as being critical to
large-scale NDSs, whose performance and effective operation require
appropriate and sufficient information exchanges among the different parts
constituting the
system~\cite{hespanha07:_survey_recen_resul_networ_contr_system,%
  baillieul07:_contr_commun_chall_networ_real_time_system}. Over the past
decade, an integrated view of information and control theories has led to new
insights into the interplay between control and communication in NDSs along
with a host of new theoretical results focusing on fundamental trade-offs
between information flow constraints and effective collective
dynamics~\cite{nair03:_expon,%
  tatikonda04:_contr,nair07:_feedb_contr_under_data_rate_const,%
  wong97:_system_finit_commun_bandw_const,%
  wong97:_system_finit_commun_bandw_const_2,%
  moreau05:_stabil_multiag_system_with_time,%
  yu10:_secon,olfati-saber07:_consen_cooper_networ_multi_agent_system,%
  hespanha07:_survey_recen_resul_networ_contr_system,%
  baillieul07:_contr_commun_chall_networ_real_time_system}.

Any information exchange---whether through a digital wireless network as in
the case of mobile sensory networks, or through a fluid as in the case of
flocking birds and schooling fish---occurs over communication channels that
are imperfect due to intrinsic limitations, primarily in terms of channel
capacity and topology. For specific classes of networked control systems,
necessary and/or sufficient conditions on the smallest data
rate~\cite{nair03:_expon,%
  tatikonda04:_contr,nair07:_feedb_contr_under_data_rate_const,%
  wong97:_system_finit_commun_bandw_const,%
  wong97:_system_finit_commun_bandw_const_2} and on the communication
topology~\cite{moreau05:_stabil_multiag_system_with_time,%
  yu10:_secon,olfati-saber07:_consen_cooper_networ_multi_agent_system,%
  hespanha07:_survey_recen_resul_networ_contr_system} for their stability or
stabilizability have been established.

Despite strong similarities between NDSs and natural swarms, there exist
numerous differences in the nature and properties of their respective
communication channels. Significant attention has been devoted to
understanding these differences in terms of topology and structure. In
particular, the emphasis has been put on the impact of dynamic and switching
topologies on the collective dynamics of locally-interacting
agents~\cite{jad,ref:ren,shang14:_influen,%
  olfati-saber07:_consen_cooper_networ_multi_agent_system}. This effort is
justified by our incomplete understanding of natural swarming behaviors, and
also by the ongoing development of new biologically-inspired designs of
artificially swarming systems. Comparatively, the specificities of the
transmission channels of naturally swarming systems in terms of capacity have
been relatively overlooked. However, the problem of reduced information flow
due to limited data rates in social transmission of information is as critical
in the natural realm as it is in the engineering one. For instance, it has
been found that the collective synchronization of neurons in dorsal root
ganglions is thwarted by a chemically-induced reduction of the firing
frequency~\cite{scholz98:_compl_block_ttx_resis_na}, which effectively
corresponds to a reduction in collective information flow associated with
smaller data rates.

Here, we investigate this problem of reduced information flow due to limited
data rates in the particular case of collective behaviors originating from
local interactions and for which information flows through a directed and
temporally-adaptive signaling network.  First, using our prior knowledge of
this adaptive interaction network and by invoking the min-flow max-cut
theorem, we identify and formalize the different possible origins of
information flow bottlenecks. We then focus on the problem of ensuring a
coherent swarming behavior and establish mathematically a sufficient condition
on the information data rate guaranteeing the emergence of a collective
response. This condition highlights the profound connection between
information flow and topology of the signaling network.  As a last step, we
provide the first investigation of the continuous phase transition---from a
globally-ordered state to a disordered one---induced by limited data
rates. The existence of a transition line relating signaling channel bandwidth
and signal-to-noise ratio is revealed and the variations along this transition
line are in good agreement with information-theoretic predictions.

\section*{Results}

\subsection*{Informational bottlenecks in collective behaviors}

%
The mechanistic quest initiated with the self-propelled-particles (SPP) model
introduced by Vicsek~\etal~\cite{vicsek95:_novel} has recently focused on
gaining insight into exactly how information flows through a swarm, with the
ultimate goal of achieving functional predictions about collective animal
behavior~\cite{strandburg-peshkin13:_visual,sumpter08:_infor,lemasson13:_motion}. Here,
we use a prototypical model of swarming, which is a refinement to the original
model by Vicsek~\etal~\cite{vicsek95:_novel} albeit based on a topological
interaction distance (see Methods).  In this class of models, like in real
swarms, collective decisions globally emerge from local information exchanges
associated with actual interagent interactions. At this stage, it is important
highlighting a notable difference between communication in NDSs and in natural
swarms. In the former, communication implies a deliberate transmission whereas
in the latter they are often not associated with deliberate exchanges of
information. A study of swarm dynamics benefits from a description and
representation of the true communication that follows those information
transfers~\cite{dusenbery92:_sensor_ecolog}. As already highlighted, any real
communication channel, irrespective of its topology and the nature of the
signal, has a finite informational capacity owing to its noisiness and limited
bandwidth. For natural swarms, channels may consist of chemicals
(e.g. chemotactic aggregation of amoebae, colony of ants) or of various forms
of energy such as electromagnetic waves and light, sound vibrations, pressure
or temperature. In practice, more than one channel may be operating
simultaneously as is the case with fish while schooling where both vision and
lateral line sensing are required~\cite{liao07:_review_of_fish_swimm_mechan}.

\begin{figure}[htbp]
  \centering
  \includegraphics[width=0.42\textwidth]{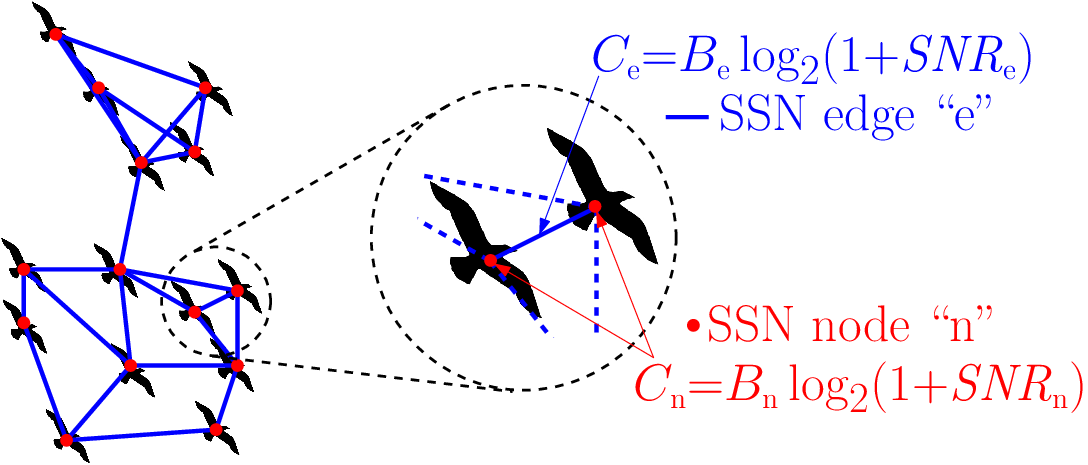}
  \caption{Schematic of a networked flock of birds with the associated
    transmission channel in the form of the swarm signaling network (SSN).
    Edges represent an interaction between two agents. Nodes are the agents
    themselves, which act as routers for the behavioral
    information.\label{fig1}}
\end{figure}

Recently, a network-theoretic approach focusing on the specificities of the
interaction network of natural swarms has emerged~\cite{%
  komareji13:_resil_contr_dynam_collec_behav,shang14:_influen,%
  strandburg-peshkin13:_visual,young13:_starl_flock_networ_manag_uncer,%
  fitch13:_infor_centr_optim_leader_selec_noisy_networ}. That approach opens
new avenues for the study of the unifying concept of swarm information flow
representing the propagation of behavioral changes. However, such a high-level
structural representation should not hide the complexity of a central part of
the real informational channel associated with the agents' sensory
cascade~\cite{dusenbery92:_sensor_ecolog}---detection, processing followed by
response---taking place when information hits a node and is routed through the
swarm signaling network
(SSN)~\cite{komareji13:_resil_contr_dynam_collec_behav}. This crucial factor
can be better fathomed when considering the prototypical swarming behavior of
predator avoidance of fish and marine
insects~\cite{radakov73:_school_ecolog_fish,treherne81:_group} or flocking
birds~\cite{attanasi14:_infor} in which the detection of an incoming predator
triggers a fright response---in the form of a swift directional change---in a
limited set of agents. These localized behavioral responses initiated by the
informed agents constitute a signal transmitted through the medium (edges of
the SSN) which, in turn, is detected by the agents (nodes of the SSN) linked
to the informed nodes (see Fig.~\ref{fig1}) if the SSN has the required
connectedness. This latter property of the SSN has been shown to primarily
depend on the interaction distance---metric, topological, or
hybrid~\cite{shang14:_consen}---and the density of swarming agents, and to
profoundly influence the consensus reaching
dynamics~\cite{shang14:_influen}. For swarms of topologically-interacting
agents, we have established that the SSN is a homogeneous, small-world and
moderately clustered
network~\cite{komareji13:_resil_contr_dynam_collec_behav}. Moreover, the SSN
is a temporal adaptive network~\cite{holme12:_tempor}, whose dynamics is
tightly coupled to that of the agents in the physical space. It is therefore
necessary to account for some of the functional details of each agent, and in
particular, its sensory cascade~\cite{dusenbery92:_sensor_ecolog}, which can
conceptually be modeled using the control-theoretic concept of multi-input and
multi-output
plant~\cite{baillieul07:_contr_commun_chall_networ_real_time_system}.

A full description of the information flow through a dynamic swarm would not
be complete without another conceptual layer borrowing elements from
information theory. As already mentioned, this integrated view of information
and control theories has been successful in advancing our understanding of the
interplay between control and communication in NDSs~\cite{nair03:_expon,%
  tatikonda04:_contr,nair07:_feedb_contr_under_data_rate_const,%
  wong97:_system_finit_commun_bandw_const,%
  wong97:_system_finit_commun_bandw_const_2,%
  moreau05:_stabil_multiag_system_with_time,%
  yu10:_secon,olfati-saber07:_consen_cooper_networ_multi_agent_system,%
  hespanha07:_survey_recen_resul_networ_contr_system,%
  baillieul07:_contr_commun_chall_networ_real_time_system}. If the total power
in a channel is distributed between the signal $S$ and random (Gaussian) noise
$N$, the Shannon--Hartley theorem provides the maximum rate of information
transmission, or capacity $C$, of the channel
\begin{equation}
  C=B
  \log_2(1+\snr),\label{eq:C}
\end{equation}
where $B$ represents the channel's bandwidth and $\snr=S/N$ is the
signal-to-noise
ratio~\cite{mackay03:_infor_theor_infer_learn_algor}. Bandwidth $B$ is
technically defined as the range of frequencies that can be transmitted, but
in natural systems the lowest frequency is always zero and it is therefore
identified with the highest frequency at which the channel can be varied. In
most real channels, the value of the bandwidth cannot always be easily
determined. In the particular case of human information exchange through
speech, it has been shown that a bandwidth of 7~kHz is required to
distinctively understand individual
syllables~\cite{meyer72:_physic_applied_acous}. In the case of swarms, one has
to consider $\Ce$ associated with informational signaling through the
medium---the network edges ``e'': e.g. the ambient fluid for flocking birds,
schooling fish and synapses for neurons---as well as $\Cn$ for the sensory
cascade internal to each agent---the network nodes ``n'', which serve as
routers for the behavioral information (see Fig.~\ref{fig1}).

Returning to the problem of global information flow through the SSN, the
max-flow min-cut theorem~\cite{papadimitriou98:_combin_optim} informs us about
informational bottlenecks, whereby the maximum information flow rate is given
by the minimum capacity of the network edges or nodes.  In other words, the
spread of information through the swarm is either limited by the signaling
through the medium (extrinsic limit) or by the agent's sensory cascade
(intrinsic limit). At this point, it is worth stressing yet another
specificity of natural swarming individuals pertaining to this intrinsic limit
to deal with information flow at the agent's level regardless of the nature of
the signal and how it propagates information through the surrounding
medium. Indeed, for most animals (including humans), there is an enormous gap
between the information capacities of sensory organs and the capacity of the
central nervous system to analyze and retain
information~\cite{dusenbery92:_sensor_ecolog}. Consequently, without loss of
generality, we can associate $\Cn$ with the information processing capacity of
the agent. To fix ideas, it is interesting knowing that in the particular case
of fire ants orientating, a rough estimate of $\Cn$ is between 0.01 and 2
bits/s~\cite{wilson62:_chemic_fr}.

As can be seen from~(\ref{eq:C}), the informational capacity is either
bandwidth- or $\snr$-limited, therefore leading to only four possible distinct
informational bottlenecks:
\begin{enumerate}[(i)]
\item low signal-to-noise ratio, $\snr\us{e}$, in the medium, which often
  occurs because of high levels of ambient noise;
\item low signal-to-noise ratio, $\snr\us{n}$, within each agent;
\item low bandwidth or data rate, $B\us{e}$, of the medium;
\item low bandwidth or data rate, $B\us{n}$, in processing information at the
  agent's central nervous system level.
\end{enumerate}
Note that option (iii) is physically irrelevant since in most media, $B\us{e}$
is typically very high~\cite{dusenbery92:_sensor_ecolog}, unless it is
artificially constrained.

\subsection*{Conditions for the emergence of collective behavior under data
  rate limitations}

%
In the sequel, we focus on the overlooked option (iv) (see Discussion) and
study the effects of information flow breakdowns in swarms stemming from
either the finiteness of the agents' bandwidth $\Bn$ or an
artificially-induced reduction in $\Bn$. Given the Nyquist theorem, to
consider the effects of a reduction in $\Bn$ is equivalent to considering an
increase in the unit interval $\Tn=1/(2B\us{n})$, which is the minimum time
interval between condition changes of data transmission signal, a.k.a. the
symbol duration time~\cite{mackay03:_infor_theor_infer_learn_algor} (see
Methods). Note that our analysis and the associated results would still hold
if we were to artificially reduce the bandwidth of the medium,
$B\us{e}=1/(2B\us{e})$ as is later suggested.

In our minimalistic model, the $N$ topologically interacting SPPs perform a
collective behavior of the consensus type for their direction of travel
$\theta_i$. At each instant, the collective state of the swarm is
characterized by $\mathbf{\Theta}(t)=\left[\theta_1(t),\ \theta_2(t),\
  \cdots,\ \theta_N(t)\right]^T$, which is updated according to the time
update rule~\eqref{swarmdisc} at the agent level (see Methods). To allow for
an analytical study of this system, we first neglect the effects of noise, and
given any initial state $\mathbf{\Theta}_0$ at time $t=0$, at any point in
time the swarm's state is
\begin{equation}
  \mathbf{\Theta}(t+m\Tn)=\left[\Pi_{j=1,\cdots,m} \Pn((j-1)\Tn) \right]
  \mathbf{\Theta}_0,
\end{equation}
given the dynamical swarm update
rule~\eqref{eq:system}. $\Pn(t)=(\mathbf{I}-\Tn\tilde{\mathbf{L}}(t))$ are
Perron matrices dependent on the unit interval
$\Tn$~\cite{olfati-saber07:_consen_cooper_networ_multi_agent_system}, with
$\tilde{\mathbf{L}}(t)=L(t)/k$, $L(t)$ being the outdegree graph Laplacian for
the SSN characterizing the instantaneous communication topology between
individuals, and $k$ is the fixed number of topological neighbors (see
Methods). It is worth adding that these Perron matrices fully embody the
instantaneous relationship between information flow and communication
structure at the core of our problem.

In the presence of a static communication topology, the stability of the
dynamical system would be governed and controlled by the spectral properties
of the constant Perron
matrix~\cite{olfati-saber07:_consen_cooper_networ_multi_agent_system}.  In the
present case, however, the constantly reconfigurable and switching network
requires a generalization of such stability analysis accounting for possibly
varying symbol duration times $\Tn$.

\bigskip
\noindent\textbf{Theorem.} \quad \itshape Let us consider the time-dependent
networked sampled-data system
\begin{equation}\label{eq:system_no_noise}
  \mathbf{\Theta}(t+\Tn)=\Pn(t)\mathbf{\Theta}(t)=(\mathbf{I}-\Tn\tilde{\mathbf{L}}(t))\mathbf{\Theta}(t),
\end{equation}
where $\tilde{\mathbf{L}}(t)=L(t)/k$, $L(t)$ being the outdegree graph
Laplacian of the network connectivity.\\
A necessary and sufficient condition for this system to be stable, and
asymptotically stable, is that it is stable at every point in time $t_j=j\Tn$.
\bigskip\normalfont

This key result is obtained by studying the convergence of infinite products
of matrices $\Pn(t_j)$ by means of the joint spectral radius $\tilde{\rho}$
defined as~\cite{berger92:_bound}
\begin{equation}
  \tilde{\rho}\mathrel{\mathop:}=\limsup_{j\rightarrow\infty}\left(\max_{t_1,\cdots,t_j\in[1,m]}\|\Pn(t_1)\cdots
    \Pn(t_j)\|\right)^{1/j}.\label{z1}
\end{equation}

\bigskip
\noindent\textbf{Proof.}
Let $\rho(\cdot)$ be the spectral radius of a matrix. By taking
$t_1=\cdots=t_j$ in Eq.~(\ref{z1}) and invoking Gelfand's spectral radius
formula, we have
\begin{equation}
  \tilde{\rho}\ge\lim_{j\rightarrow\infty}\|\Pn(t)^j\|^{1/j}=\rho(\Pn(t)),\label{z2}
\end{equation}
for any $t\in[1,m]$. Therefore, if $\rho(\Pn(t))>1$, for any $t$, then
$\tilde{\rho}>1$. On the other hand, for any $\varepsilon>0$, there exists a
matrix norm $\|\cdot\|$ such that (e.g.~\cite[Lemma
5.6.10]{horn87:_matrix_analy})
\begin{eqnarray}
  \tilde{\rho}&\le&\lim_{j\rightarrow\infty}\left(\max_{t_1,\cdots,t_j\in[1,m]}\|\Pn(t_1)\|^{1/j}\right)\cdots\nonumber\\
  &&\cdot
  \left(\max_{t_1,\cdots,t_j\in[1,m]}\|\Pn(t_j)\|^{1/j}\right)\nonumber\\
  &=&\max_{t\in[1,m]}\|\Pn(t)\|\nonumber\\
  &\le&\max_{t\in[1,m]}\{\rho(\Pn(t))+\varepsilon\}.\label{z3}
\end{eqnarray}
Therefore, if $\rho(\Pn(t))<1$ for all $t$, we can choose $\varepsilon$ small
enough so that $\tilde{\rho}<1$. Recall that the system
(\ref{eq:system_no_noise}) is
stable if and only if $\tilde{\rho}<1$~\cite{berger92:_bound,jad}.\\
$\Box$ \bigskip

The above Theorem leads to the following Corollary, which has very important
practical implications, as it guarantees the system stability, namely, the
consensus reaching of the swarm under certain conditions on the symbol
duration $\Tn$.

\bigskip
\noindent\textbf{Corollary.} \quad \itshape
A sufficient condition for the stability of the networked sampled-data
system~\eqref{eq:system_no_noise} is given by the following upper bound on the
symbol duration time, which has to be verified at every single point in time
$t_j=j\Tn$:
\begin{equation}
  \Tn<\frac{2}{\max_{1\le i\le
      N}|\lambda_i(\tilde{\mathbf{L}}(t))|}\quad \mathrm{for}\
  \mathrm{all}\ t,\label{z}
\end{equation}
where $\lambda_i(\tilde{\mathbf{L}}(t))$ are the eigenvalues of
$\tilde{\mathbf{L}}(t)=L(t)/k$, $L(t)$ being the outdegree graph
Laplacian of the network connectivity.\\
\bigskip\normalfont

\noindent\textbf{Proof.}
It follows from (\ref{z}) that, for all $i$ and $t$,
$0\le\lambda_i(\Tn\tilde{\mathbf{L}}(t))<2$, and hence
$-1<\lambda_i(\Pn(t))\le1$. Note that we are unable to sharpen the upper bound
to $\lambda_i(\Pn(t))<1$ (hence, we cannot conclude the stability immediately)
since 0 is always an eigenvalue of $\Tn\tilde{\mathbf{L}}(t)$ for any
$\Tn\ge0$. However, we know that as $\Tn\rightarrow0$ the corresponding
continuous system is stable.  Therefore, when $\Tn$ becomes small enough (as
specified by (\ref{z})), our system (\ref{eq:system_no_noise}) is also stable
and $\tilde{\rho}<1$ follows. The above Theorem therefore allows us to
conclude
the proof of this Corollary.\\
$\Box$ \bigskip

Based on condition (\ref{z}) and the relation $\Tn=1/(2B\us{n})$ between unit
interval and bandwidth, we find that if the bandwidth satisfies the sufficient
condition
\begin{equation}
  \Bn > \Bn^0=\frac{1}{4}\max_{1\le i\le N}|\lambda_i(\tilde{\mathbf{L}}(t))|\quad
  \mathrm{for}\ \mathrm{all}\ t,\label{eq:lower-bound}
\end{equation}
then the consensus reaching of the swarm is guaranteed. Note that the
superscript ``0'' in $\Bn^0$ serves as reminder that this analytic derivation
was obtained in the absence of intrinsic or extrinsic sources of noise.

\subsection*{Analysis of swarming collapse under data rate limitations}
%

\begin{figure}[htbp]
  \centering
  \begin{overpic}[width=.34\textwidth]{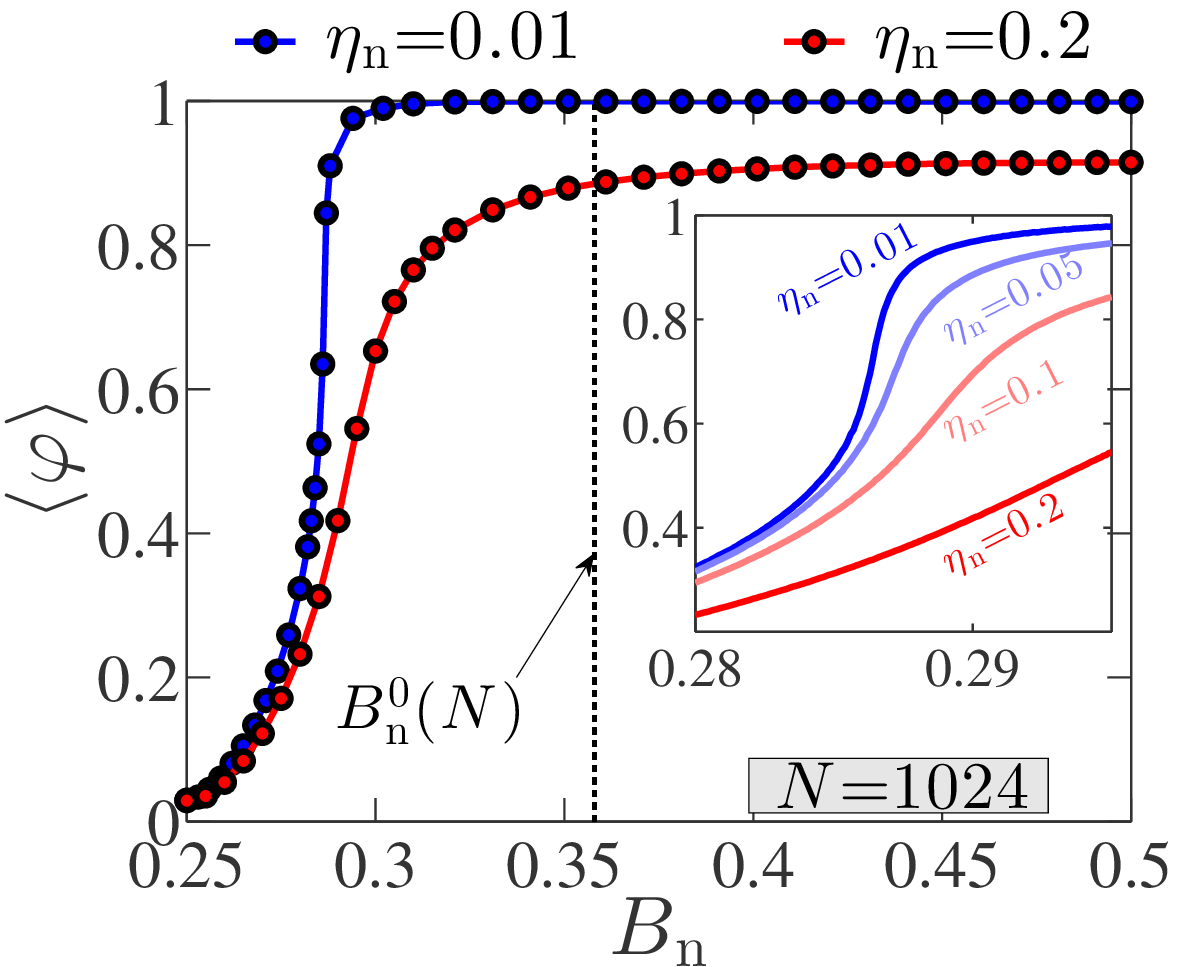}
    \put(0,74){(a)}
  \end{overpic}
  \\[1ex]
  \centering
  \begin{overpic}[width=.34\textwidth]{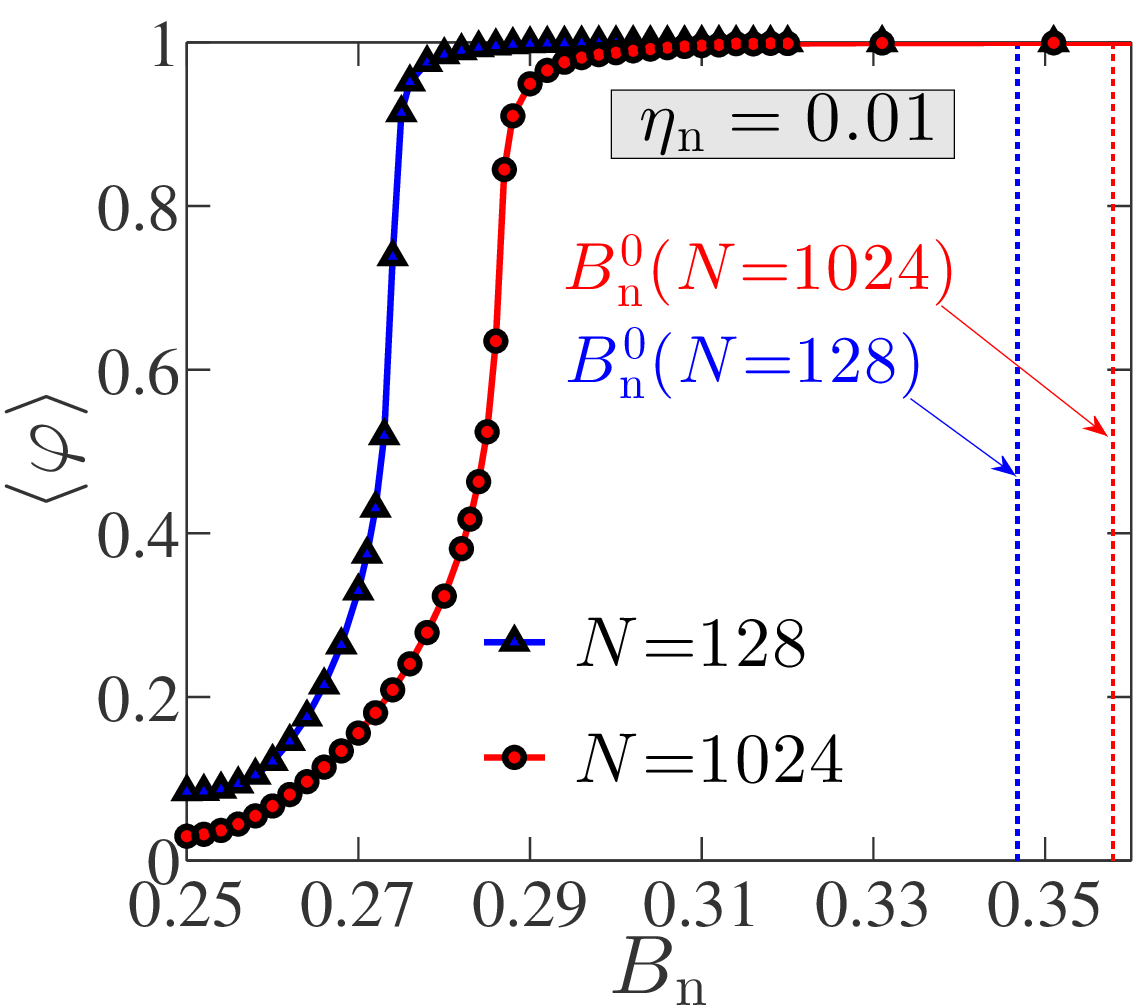}
    \put(0,82){(b)}
  \end{overpic}
  \begin{overpic}[width=.34\textwidth]{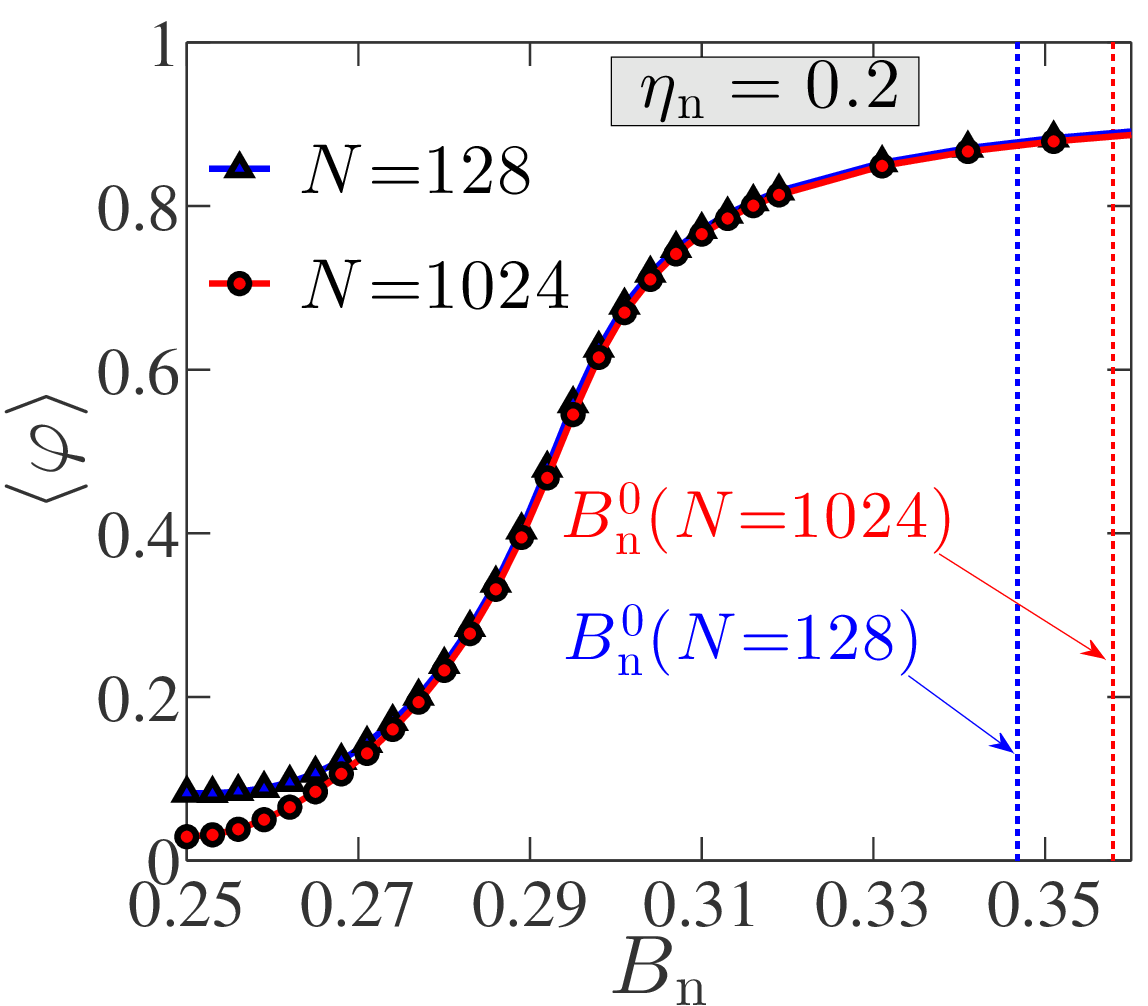}
    \put(0,82){(c)}
  \end{overpic}
  \caption{ Collapse of swarming with decreasing $\Bn$: (a) $N{=}1024$; (b)
    $\eta\us{n}{=}1\%$; (c) $\eta\us{n}{=}20\%$. Values for $\Bn^0(N)$ are
    obtained from~(\ref{eq:lower-bound}) in the $\eta\us{n}{=}0$ limit upon
    averaging over a sample of $10^4$ SSNs. ($v_0{=}0.3$, $k{=}7$,
    $\rho{=}N/\ell^2{=}100$, and equivalent statistics for all data
    points.)\label{fig2}}
\end{figure}

%
As a next step, we seek evidences of such required minimum information flow by
simulating the dynamics of $N$ SPPs governed by~(\ref{swarmdisc}), with
decreasing bandwidth $\Bn$---the control parameter---in the presence of
different levels of intrinsic noise $\eta\us{n}$ (see Methods). The
effectiveness in swarming is classically measured by the order parameter $
\varphi(t)\equiv \frac{1}{N} \left|\sum^{N}_{j = 1}\textrm{e}^{\textrm{i}
    \theta_j(t) }\right|$. For large bandwidths, $\Bn\gg \Bn^0$, swarms of
vastly different sizes systematically produce large-scale order, even in the
presence of relatively high noise levels (see Fig.~\ref{fig2}). Continued
reduction in $\Bn$ below $\Bn^0$ consistently yields a swarming
collapse---corresponding to a disordered state of the system lacking
large-scale self-organization---irrespective of the swarm size $N$ or noise
level $\eta\us{n}$ (see Fig.~\ref{fig2}).

\begin{figure}[htbp]
  \centering
  \begin{overpic}[width=.34\textwidth]{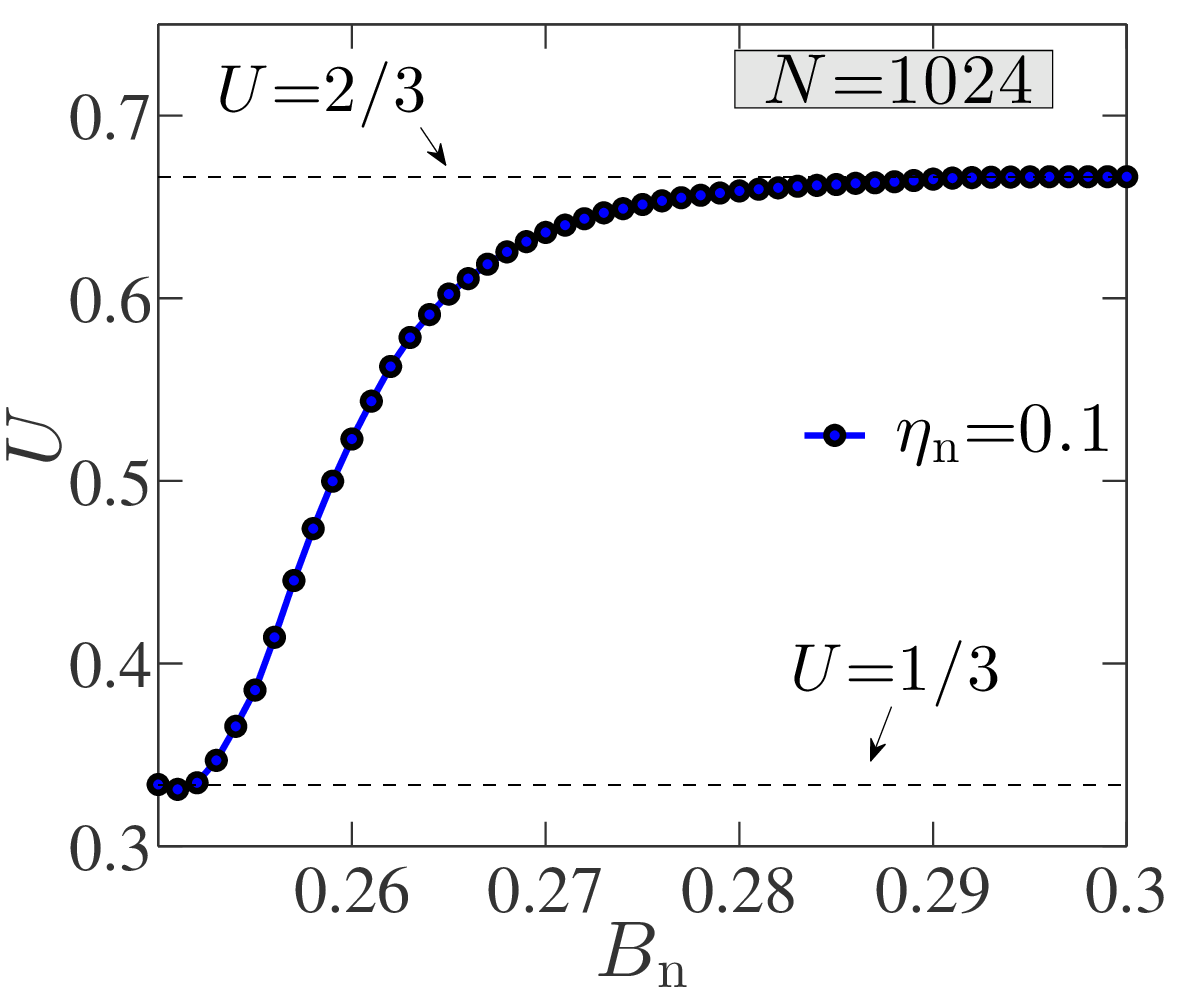}
    \put(0,82){(a)}
  \end{overpic}
  \begin{overpic}[width=.34\textwidth]{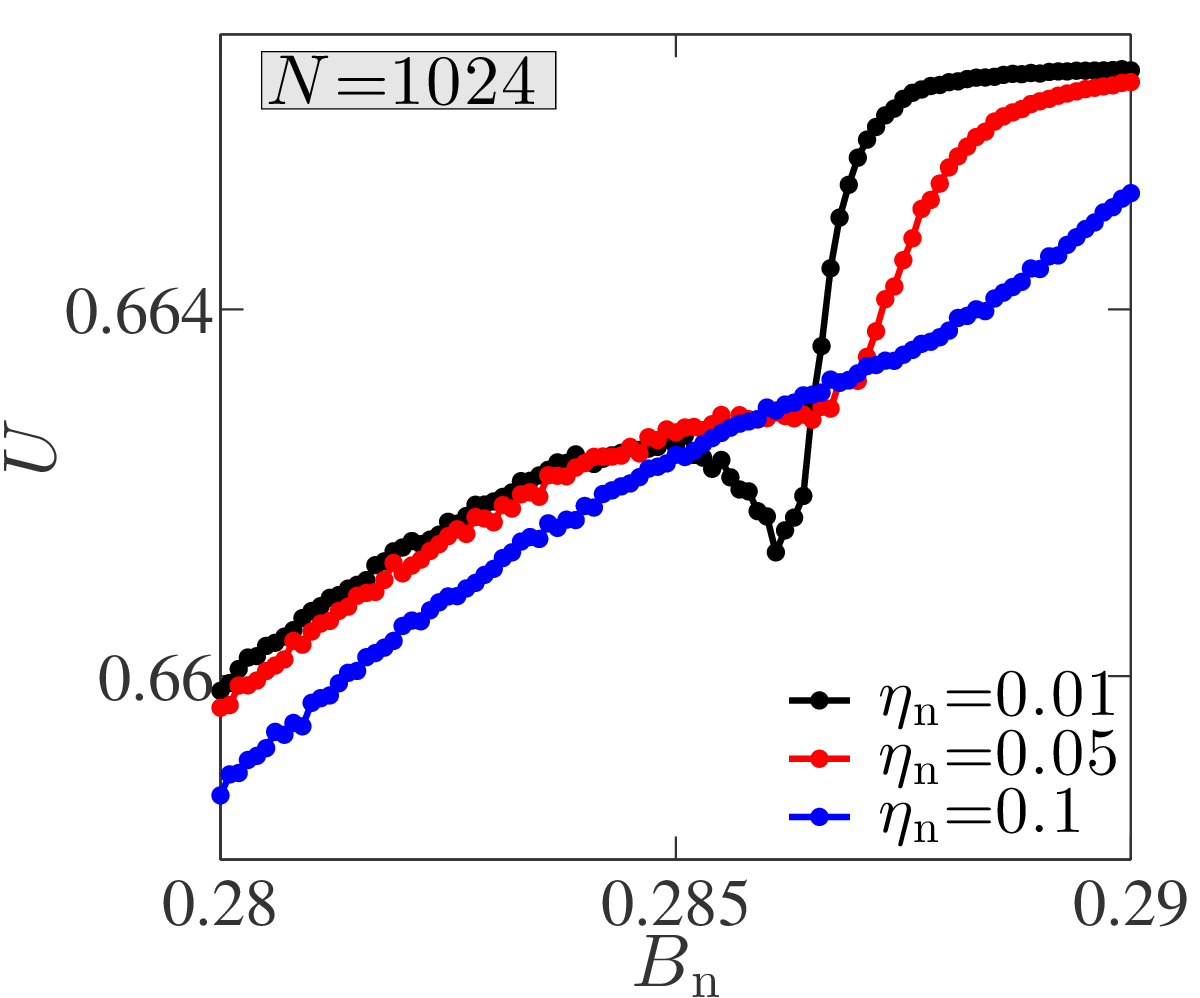}
    \put(0,82){(b)}
  \end{overpic}
  \caption{Stationary values of $U$: (a) wide $\Bn$ range; (b) At the critical
    bandwidth $\Bn^c{=}0.286\pm 0.001$, $U$ has the same value for not too
    high $\eta\us{n}$. ($v_0{=}0.3$, $k{=}7$, $\rho{=}N/\ell^2{=}100$, and
    equivalent statistics for all data points.)\label{fig3}}
\end{figure}

These phase transitions are of second order since the Binder
cumulant~\cite{binder81:_finit_size_scalin_analy_ising}, $U\equiv 1 - \langle
\varphi^4 \rangle/3\langle \varphi^2 \rangle^2$, remains positive for all
values of the control parameter $\Bn$ (see Fig.~\ref{fig3}(a)). However, it is
very likely that similarly to noise-induced phase transitions, the observation
of continuous phase transitions is only apparent owing to strong small-size
effects~\cite{gregoire04:_onset_of_collec_and_cohes_motion,%
  chate08:_collec,solon14:_from_phase_micro_phase_separ_flock_model}. Indeed,
in our particular framework, we are dealing with real-life finite-size
swarms. For such swarms, the population $N$ is relatively small, especially
compared to the thermodynamic limit, which is classically invoked to fully
characterize the very nature of a phase transition from the statistical
physics standpoint.

\begin{figure}[htbp]
  \centering
  \begin{overpic}[width=.34\textwidth]{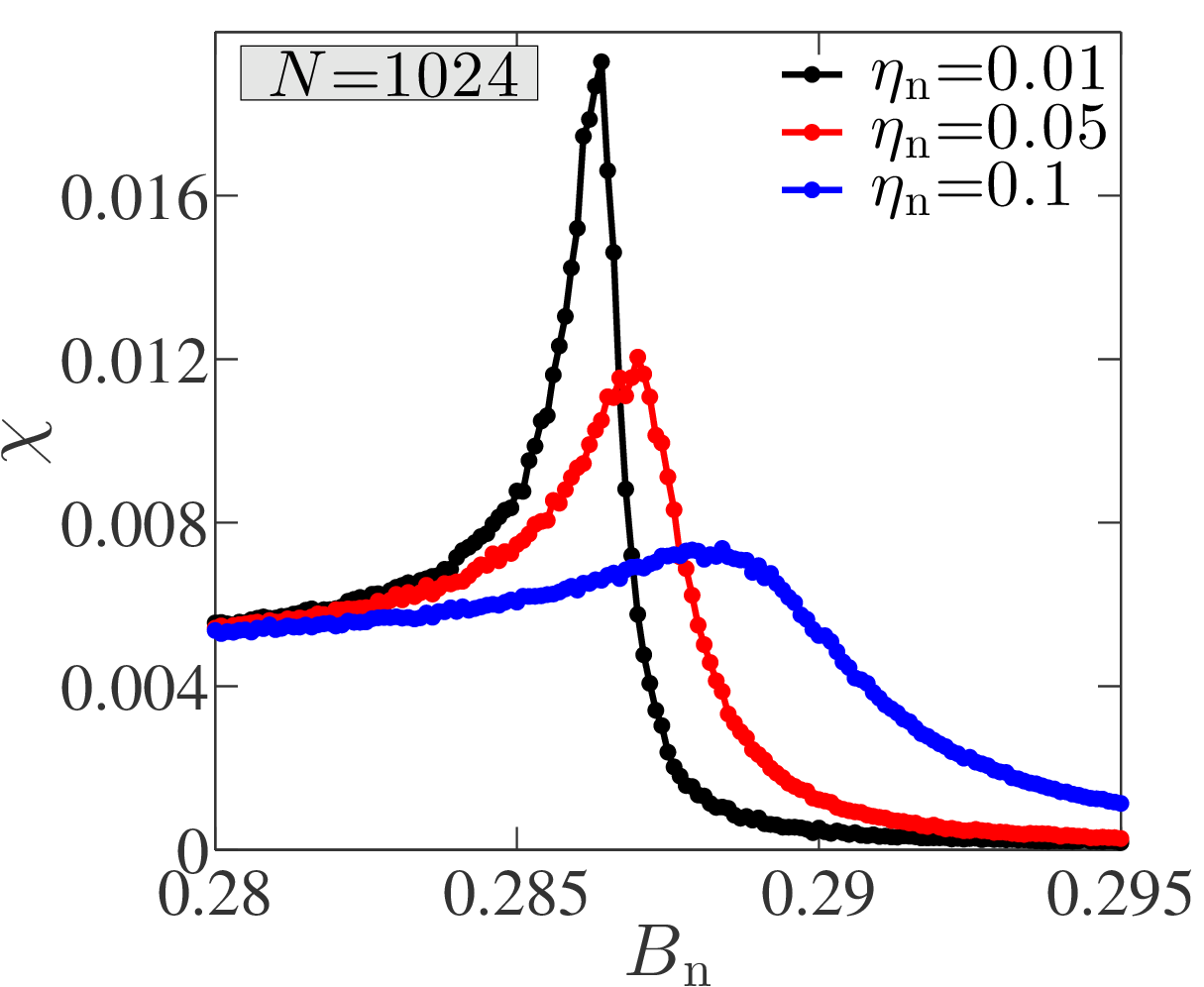}
    \put(0,82){(a)}
  \end{overpic}
  \begin{overpic}[width=.34\textwidth]{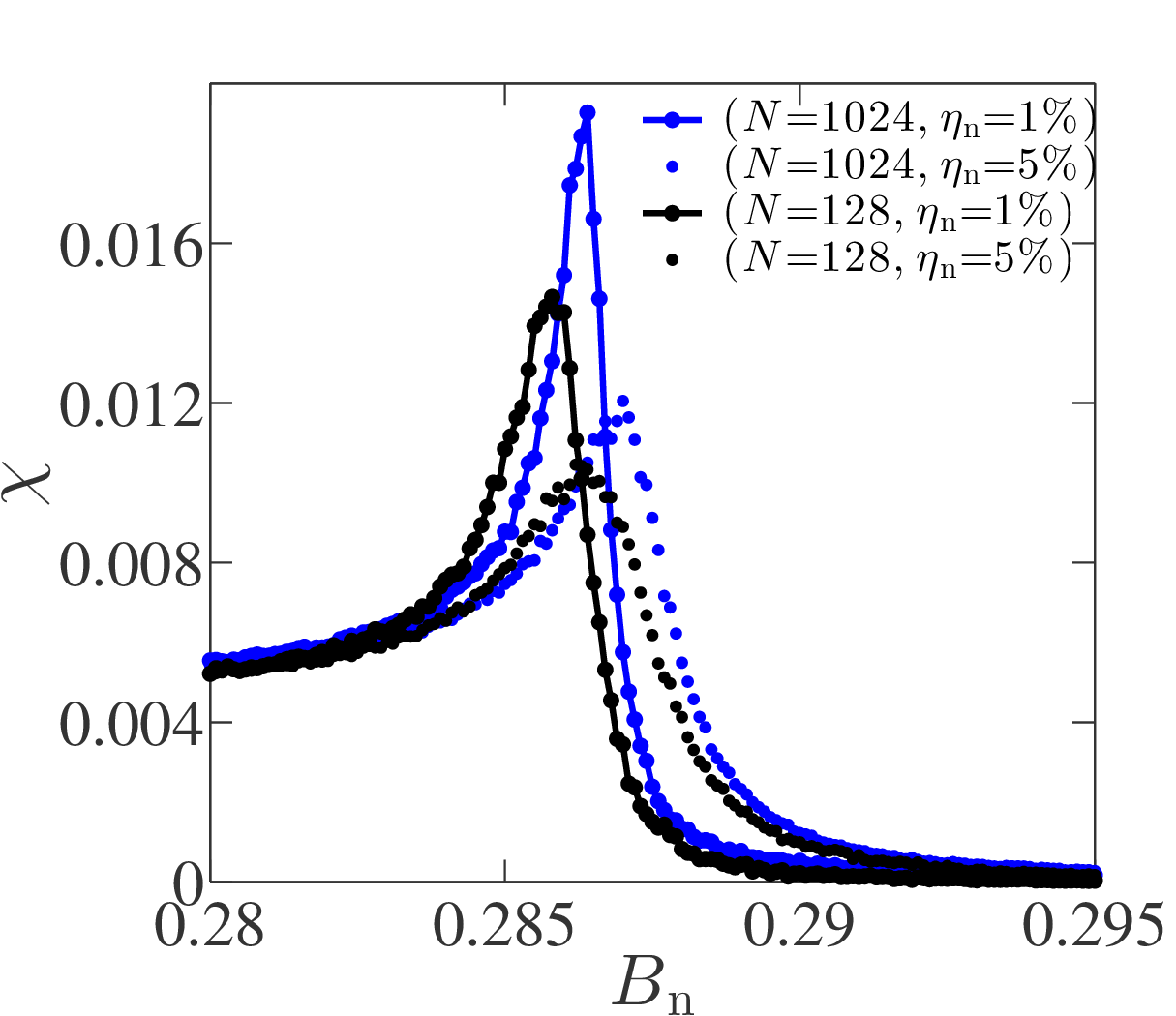}
    \put(0,82){(b)}
  \end{overpic}
  \caption{Stationary values of $\chi$: (a) $N=1024$; (b) $N=128$ and
    $N=1024$. ($v_0{=}0.3$, $k{=}7$, $\rho{=}N/\ell^2{=}100$, and equivalent
    statistics for all data points.)\label{fig4}}
\end{figure}

We further observe the existence of a transition line for which the critical
bandwidth varies with the noise, i.e. $\Bn^c=\Bn^c(\eta\us{n})$. As expected,
the variations of the susceptibility $\chi\equiv \ell^2 (\langle
\varphi^2\rangle -\langle \varphi \rangle^2)$ with $\Bn$ reveal the occurrence
of large fluctuations of the order parameter near the phase transition (see
Fig.~\ref{fig4}).
It is worth adding that all the above observations remain unchanged for larger
values of $N$, other values of the density $\rho=N/\ell^2$, for a wide range
of $v_0$, and for other values of $k>7$. As mentioned earlier, in the
thermodynamic limit, we still expect to obtain a phase transition albeit
possibly of the first order kind as is the case with noise-induced phase
transitions~\cite{gregoire04:_onset_of_collec_and_cohes_motion,chate08:_collec,solon14:_from_phase_micro_phase_separ_flock_model}. However,
our focus here was on swarms numbering in the thousand---as is typical with
topologically-interacting flocks of starlings~\cite{bialek12:_statis}, and was
not on determining the fine nature of the phase transition taking place when
reducing the bandwidth.

\section*{Discussion}
%

%
The study of the effects of reduced accuracy---owing to the ubiquitous
presence of environmental noise combined to limited sensory capabilities---in
social transmission of information has received significant attention at both
the experimental and modeling levels~\cite{viscek2012}. Comparatively, the
consequences of limited social information flow within natural swarms have
been relatively overlooked despite the realization of its significance in
networked control system theory over a decade
ago~\cite{elia01:_stabil_linear_system_with_limit_infor,%
  wong97:_system_finit_commun_bandw_const,%
  baillieul07:_contr_commun_chall_networ_real_time_system,%
  nair07:_feedb_contr_under_data_rate_const}.  Information is often defined as
being the capacity to organize a system. This definition resonates very well
with our search for informational bottlenecks hindering the self-organization
of swarms. Combining network-theoretic and information-theoretic elements, our
analysis reveals 4 possible sources of informational bottleneck: (i)--(iv)
(see Results). As already highlighted, option (iii), which corresponds to low
$B\us{e}$, is physically irrelevant for naturally evolving swarms. The
influence of low $\snr\us{e}$ in the medium, option (i), can be used to
explain empirical evidences of some specific swarming breakdowns: e.g. schools
of fish disperse at dusk~\cite{glass86:_atlan_scomb,%
  keenleyside55:_some_aspec_school_behav_fish,%
  emery73:_prelim_compar_day_night_habit}.  Essentially, the phase transitions
uncovered using Vicsek's model and its variations~\cite{viscek2012}, from a
collectively ordered phase to a disordered one following an increase in
ambient noise level, can be traced back to information flow breakdowns through
noisy channels: either with low
$\snr\us{e}$~\cite{gregoire04:_onset_of_collec_and_cohes_motion} as in option
(i), or with low $\snr\us{n}$~\cite{vicsek95:_novel} as in option (ii).  Other
empirical evidences are stressing the importance of a sufficiently-high
bandwidth~\footnote{Equivalent to information update since the Nyquist rate
  $2B=f$ relates bandwidth and frequency of update} in complex adaptive
systems: e.g. information transmission through signaling relay during the
collective migration of social amoebae~\cite{mccann10:_cell}, or the induction
of differential anesthesia when chemically reducing the firing frequency of
neurons in dorsal root
ganglions~\cite{scholz98:_compl_block_ttx_resis_na}. These empirical evidences
are additional motivations for our investigation of the overlooked option (iv)
and the associated study of the effects of information flow breakdowns in
swarms stemming from either the finiteness of the agents' bandwidth $\Bn$ or
an artificially-induced reduction in $\Bn$ or $B\us{e}$.

The study of these limiting effects on swarm dynamics has been carried out
using two complementary approaches. First, we neglected the effects of noise
in Eq.~\eqref{eq:system} and modeled the swarm as a networked sampled-data
system governed by Eq.~\eqref{eq:system_no_noise} (see Methods \&
Results). The study of the stability and asymptotic stability of the swarm
dynamics led to Eq.~(\ref{eq:lower-bound}). This sufficient condition on the
bandwidth $\Bn$ reveals the profound connection between, on the one hand the
switching communication topology---through the maximum eigenvalue of the
normalized directed graph Laplacian $\tilde{\mathbf{L}}(t)$ of the signaling
network, and on the other hand, the necessary information flow for effective
swarming measured by the bandwidth $\Bn$. We speculate that the effects of not
satisfying this condition (i.e. having $\Bn>\Bn^0$) can readily be tested
experimentally using a simple setup consisting of fish schooling in a tank
with a stroboscopic light shining onto them. By reducing the frequency $f$ of
the flash, we artificially force the decrease in $\Bn$ (and $B\us{e}$) and we
expect that at a given critical frequency $f^c=2\Bn^c$, the coordinated
schooling behavior will disappear. We further speculate that owing to their
evolutionary-optimized character, biological swarming agents such as fish and
birds do not naturally exhibit the bandwidth-induced swarming collapse
uncovered here. Beyond the consequences for self-organized biological systems,
our work also highlights the importance of having sufficient information
signaling capacity when designing artificial swarms so as to ensure their
effectiveness. At a more qualitative level, our analysis also provides the
first physical explanation for the required minimum firing frequency of
neurons in dorsal root ganglions to maintain consciousness and thereby avoid
inducing differential anesthesia~\cite{scholz98:_compl_block_ttx_resis_na}.

The second approach used to study the limiting effects of information
bottlenecks on swarm dynamics consists in carrying out actual SPP
simulations. That approach has the advantage of incorporating the combined
effects of $\snr\us{n}$ and $\Bn$. More importantly, it allows us to carefully
study changes in the swarm dynamics in the vicinity of the collapse of
long-range order, i.e. the phase transition. That was not possible with our
first approach that focused on preventing a swarming collapse in the absence
of noise. It is worth adding that most SPP simulations heretofore reported in
the literature (e.g.~\cite{lemasson13:_motion,
  komareji13:_resil_contr_dynam_collec_behav,bialek12:_statis,
  gregoire04:_onset_of_collec_and_cohes_motion,shang14:_influen}) have been
generated using arbitrary, yet sufficiently high, bandwidth levels. Therefore,
in those past works, the collapse of swarming is rooted in the noisiness of
the signaling channel. For all swarm sizes, the simulation results confirm the
occurrence of a swarming collapse as expected from our analytical study of the
networked sampled-data system~\eqref{eq:system_no_noise} (see Results). The
key point revealed by these simulations concerns the existence of a transition
line for which the critical bandwidth varies with the noise,
i.e. $\Bn^c=\Bn^c(\eta\us{n})$. The existence of this transition line could
have been anticipated from the interplay between noise and bandwidth
originating from the Shannon--Hartley theorem and the expression~\eqref{eq:C}
for the channel capacity. Indeed, along the transition line
$\Bn^c(\eta\us{n})$, we have that $\Bn^c$ decreases with decreasing
$\eta\us{n}$. This important observation is consistent with our intuition that
a higher noise level would require a higher volume of information to be
exchanged for the swarm to self organize. From the information-theoretic
viewpoint, this trend can readily be explained if we assume the existence of a
minimum ``critical'' rate of information $R^c$ below which a collapse of
swarming occurs. At the critical point, the max-flow min-cut
theorem~\cite{papadimitriou98:_combin_optim} gives us $\Cn=R^c$ and given
expression~(\ref{eq:C}), we have that $\Bn^c\downarrow$ with
$\eta\us{n}\downarrow$. Despite the singularity in Shannon's capacity at the
zero-noise limit, our approach allows us to determine the critical bandwidth
in this limit through the intersection of the $U(\Bn)$ for several nonzero
values of $\eta\us{n}$ (see Fig.~\ref{fig3}(b)).

\section*{Methods}

%
We consider a minimalists model for the swarming system, which consists of $N$
topologically interacting
SPPs~\cite{vicsek95:_novel,strandburg-peshkin13:_visual,%
  komareji13:_resil_contr_dynam_collec_behav,%
  young13:_starl_flock_networ_manag_uncer,bialek12:_statis,shang14:_influen},
moving at constant speed $v_0$ through a $\ell \times \ell$ domain having
periodic boundaries.  Each individual $i$ is characterized by its direction of
travel $\theta_i$, and a canonical swarming behavior of the consensus type is
examined. To account for the finiteness of the bandwidth, we consider
synchronous information exchanges occuring every $\Tn=1/(2B\us{n})$, where the
unit interval $\Tn$ is the minimum time interval between condition changes of
data transmission signal, a.k.a. the symbol duration
time~\cite{mackay03:_infor_theor_infer_learn_algor}. The agents move
synchronously at discrete time steps $\Tn$ by a fixed distance $\delta=v_0\Tn$
upon receiving informational signals from their neighbors as per the linear
update rule
\begin{equation}\label{swarmdisc}
  \theta_i(t+\Tn) =\theta_i(t)+\frac{\Tn}{k_i} \sum_{j \sim i}
  \left\{ \theta_j(t)-\theta_i(t)\right\} + \eta\us{n} \xi_i(t),
\end{equation}
where $k_i=k$ is the fixed number of individuals in the topological
neighborhood $j\sim i$ of $i$ and $\eta\us{n}\xi_i(t)$ is a Gaussian white
noise ($\xi_i(t) \in [-\pi,\pi]$).

Even though directions are intrinsically nonlinear quantities, such linear
consensus models are
known~\cite{olfati-saber07:_consen_cooper_networ_multi_agent_system,%
  jad,young13:_starl_flock_networ_manag_uncer,komareji13:_resil_contr_dynam_collec_behav,shang14:_influen}
to yield phase transitions similar to those obtained with nonlinear models,
such as the one in Ref.~\cite{vicsek95:_novel}. In addition,
Eq.~(\ref{swarmdisc}) is a discrete-time version of the minimal model
consistent with experimental correlations in natural flocks of birds, while
also predicting the propagation of order throughout entire
flocks~\cite{bialek12:_statis,young13:_starl_flock_networ_manag_uncer}. Note
that the growing body of evidence in support of a topological model of
interaction between flocking birds~\cite{strandburg-peshkin13:_visual,%
  komareji13:_resil_contr_dynam_collec_behav,%
  young13:_starl_flock_networ_manag_uncer,bialek12:_statis,shang14:_influen}
guided our choice. However, qualitatively similar results were obtained with
the exact same model with metric interactions, which is consistent with the
recent evidence of a unique universality class in the noise-induced critical
behavior of SPPs, regardless of the metric or topological nature of
interactions~\cite{barberis14:_eviden}.

At each instant, the dynamical swarm behavior is governed by:
\begin{equation}\label{eq:system}
  \mathbf{\Theta}(t+\Tn)=\Pn(t)\mathbf{\Theta}(t)+\eta\us{n} \mathbf{\Xi}(t),
\end{equation}
with ($\mathbf{\Theta}(t),\mathbf{\Xi}(t))=(\{\theta_i(t)\}_{1\leq i \leq
  N}^{\textrm{T}},\{\xi_i(t)\}_{1\leq i \leq N}^{\textrm{T}})$ and
$\Pn(t)=(\mathbf{I}-\Tn\tilde{\mathbf{L}}(t))$ are Perron matrices dependent
on the unit interval
$\Tn$~\cite{olfati-saber07:_consen_cooper_networ_multi_agent_system}, with
$\tilde{\mathbf{L}}(t)=L(t)/k$, $L(t)$ being the outdegree graph Laplacian for
SSN characterizing the instantaneous communication topology between
individuals. The system~(\ref{eq:system}) fully embodies the time-dependent
relationship between the information flow and communication structure at the
core of our problem.

\acknowledgements This work was supported by a grant from the Temasek Lab
(TL@SUTD) under the STARS project, by a grant from the SUTD-MIT International
Design Centre, and by a grant from the Singapore-MIT Alliance for Research and
Technology (SMART) under the Pilot Phase II program.

\section*{References}
%

\begin{thebibliography}{10}
  \expandafter\ifx\csname url\endcsname\relax \def\url#1{\texttt{#1}}\fi
  \expandafter\ifx\csname urlprefix\endcsname\relax\def\urlprefix{URL }\fi
  \providecommand{\bibinfo}[2]{#2} \providecommand{\eprint}[2][]{\url{#2}}

\bibitem{bouffanais} \bibinfo{author}{Bouffanais, R.}  \newblock
  \emph{\bibinfo{title}{Design and Control of Swarm Dynamics}}.  \newblock
 Complexity Series
  (\bibinfo{publisher}{Springer}, \bibinfo{address}{Singapore}, \bibinfo{year}{2016}).

\bibitem{sumpter06:_princ_of_collec_animal_behav} \bibinfo{author}{Sumpter,
    D. J.~T.}  \newblock \bibinfo{title}{The principles of collective animal
    behaviour}.  \newblock \emph{\bibinfo{journal}{Phil. Trans. R. Soc. B}}
  \textbf{\bibinfo{volume}{361}}, \bibinfo{pages}{5--22}
  (\bibinfo{year}{2006}).

\bibitem{mccann10:_cell} \bibinfo{author}{McCann, C.},
  \bibinfo{author}{Kriebel, P.}, \bibinfo{author}{Parent, C.} \&
  \bibinfo{author}{Losert, W.}  \newblock \bibinfo{title}{Cell speed,
    persistence and information transmission during signal relay and
    collective migration}.  \newblock \emph{\bibinfo{journal}{J. Cell Sci.}}
  \textbf{\bibinfo{volume}{123}}, \bibinfo{pages}{1724--1731}
  (\bibinfo{year}{2010}).

\bibitem{hsieh08:_decen} \bibinfo{author}{Hsieh, M.~A.},
  \bibinfo{author}{Kumar, V.} \& \bibinfo{author}{Chaimowicz, L.}  \newblock
  \bibinfo{title}{Decentralized controllers for shape generation with robotic
    swarms}.  \newblock \emph{\bibinfo{journal}{Robotica}}
  \textbf{\bibinfo{volume}{26}}, \bibinfo{pages}{691--701}
  (\bibinfo{year}{2008}).

\bibitem{bialek12:_statis} \bibinfo{author}{Bialek, W.} \emph{et~al.}
  \newblock \bibinfo{title}{Statistical mechanics for natural flocks of
    birds}.  \newblock \emph{\bibinfo{journal}{Proc. Natl. Acad. Sci. {USA}}}
  \textbf{\bibinfo{volume}{109}}, \bibinfo{pages}{4786--4791}
  (\bibinfo{year}{2012}).

\bibitem{attanasi14:_infor} \bibinfo{author}{Attanasi, A.} \emph{et~al.}
  \newblock \bibinfo{title}{Information transfer and behavioural inertia in
    starling flocks}.  \newblock \emph{\bibinfo{journal}{Nature Phys.}}
  \textbf{\bibinfo{volume}{10}}, \bibinfo{pages}{691--696}
  (\bibinfo{year}{2014}).

\bibitem{bazazi08:_collec} \bibinfo{author}{Bazazi, S.} \emph{et~al.}
  \newblock \bibinfo{title}{Collective motion and cannibalism in locust
    migratory bands}.  \newblock \emph{\bibinfo{journal}{Curr. Biol.}}
  \textbf{\bibinfo{volume}{18}}, \bibinfo{pages}{735--739}
  (\bibinfo{year}{2008}).

\bibitem{shang14:_influen} \bibinfo{author}{Shang, Y.} \&
  \bibinfo{author}{Bouffanais, R.}  \newblock \bibinfo{title}{Influence of the
    number of topologically interacting neighbors on swarm dynamics}.
  \newblock \emph{\bibinfo{journal}{Sci. Rep.}} \textbf{\bibinfo{volume}{4}},
  \bibinfo{pages}{4184} (\bibinfo{year}{2014}).

\bibitem{haque11:_biolog} \bibinfo{author}{Haque, M.~A.},
  \bibinfo{author}{Rahmani, A.~R.} \& \bibinfo{author}{Egerstedt, M.~B.}
  \newblock \bibinfo{title}{Biologically inspired confinement of multi-robot
    systems}.  \newblock \emph{\bibinfo{journal}{Int. J. Bio-Inspired
      Comput.}}  \textbf{\bibinfo{volume}{3}}, \bibinfo{pages}{213--224}
  (\bibinfo{year}{2011}).

\bibitem{krause02:_livin_in_group} \bibinfo{author}{Krause, J.} \&
  \bibinfo{author}{Ruxton, G.~D.}  \newblock \emph{\bibinfo{title}{Living in
      Groups}}.  \newblock Oxford Series in Ecology and Evolution
  (\bibinfo{publisher}{Oxford University Press}, \bibinfo{address}{Oxford,
    U.K.}, \bibinfo{year}{2002}).

\bibitem{sumpter10:_collec_animal_behav} \bibinfo{author}{Sumpter, D. J.~T.}
  \newblock \emph{\bibinfo{title}{Collective Animal Behavior}}
  (\bibinfo{publisher}{Princeton University Press},
  \bibinfo{address}{Princeton, NJ}, \bibinfo{year}{2010}).

\bibitem{strandburg-peshkin13:_visual} \bibinfo{author}{Strandburg-Peshkin,
    A.} \emph{et~al.}  \newblock \bibinfo{title}{Visual sensory networks and
    effective information transfer in animal groups}.  \newblock
  \emph{\bibinfo{journal}{Curr. Biol.}} \textbf{\bibinfo{volume}{23}},
  \bibinfo{pages}{R709--R711} (\bibinfo{year}{2013}).

\bibitem{sumpter08:_infor} \bibinfo{author}{Sumpter, D.},
  \bibinfo{author}{Buhl, J.}, \bibinfo{author}{Biro, D.} \&
  \bibinfo{author}{Couzin, I.}  \newblock \bibinfo{title}{Information transfer
    in moving animal groups}.  \newblock \emph{\bibinfo{journal}{Theory
      Biosci.}}  \textbf{\bibinfo{volume}{127}}, \bibinfo{pages}{177--186}
  (\bibinfo{year}{2008}).

\bibitem{lemasson13:_motion} \bibinfo{author}{Lemasson, B.},
  \bibinfo{author}{Anderson, J.} \& \bibinfo{author}{Goodwin, R.}  \newblock
  \bibinfo{title}{Motion-guided attention promotes adaptive communications
    during social navigation}.  \newblock
  \emph{\bibinfo{journal}{Proc. R. Soc. B}} \textbf{\bibinfo{volume}{280}},
  \bibinfo{pages}{20122003} (\bibinfo{year}{2013}).

\bibitem{szabo09:_trans} \bibinfo{author}{Szab\'o, P.}, \bibinfo{author}{Nagy,
    M.} \& \bibinfo{author}{Vicsek, T.}  \newblock \bibinfo{title}{Transitions
    in a self-propelled-particles model with coupling of accelerations}.
  \newblock \emph{\bibinfo{journal}{Phys. Rev. E}}
  \textbf{\bibinfo{volume}{79}}, \bibinfo{pages}{021908}
  (\bibinfo{year}{2009}).

\bibitem{bode10:_how} \bibinfo{author}{Bode, N. W.~F.},
  \bibinfo{author}{Faria, J.~J.}, \bibinfo{author}{Franks, D.~W.},
  \bibinfo{author}{Krause, J.} \& \bibinfo{author}{Wood, A.~J.}  \newblock
  \bibinfo{title}{How perceived threat increases synchronization in
    collectively moving animal groups}.  \newblock
  \emph{\bibinfo{journal}{Proc. R. Soc. B}} \textbf{\bibinfo{volume}{277}},
  \bibinfo{pages}{3065--3070} (\bibinfo{year}{2010}).

\bibitem{handegard12:_dynam_coord_group_huntin_collec}
  \bibinfo{author}{Handegard, N.~O.} \emph{et~al.}  \newblock
  \bibinfo{title}{The dynamics of coordinated group hunting and collective
    information transfer among schooling prey}.  \newblock
  \emph{\bibinfo{journal}{Curr. Biol.}} \textbf{\bibinfo{volume}{22}},
  \bibinfo{pages}{1213--1217} (\bibinfo{year}{2012}).

\bibitem{hespanha07:_survey_recen_resul_networ_contr_system}
  \bibinfo{author}{Hespanha, J.~P.}, \bibinfo{author}{Naghshtabrizi, P.} \&
  \bibinfo{author}{Xu, Y.}  \newblock \bibinfo{title}{A survey of recent
    results in networked control systems}.  \newblock
  \emph{\bibinfo{journal}{Proc. IEEE}} \textbf{\bibinfo{volume}{95}},
  \bibinfo{pages}{138--162} (\bibinfo{year}{2007}).

\bibitem{baillieul07:_contr_commun_chall_networ_real_time_system}
  \bibinfo{author}{Baillieul, J.} \& \bibinfo{author}{Antsaklis, P.~J.}
  \newblock \bibinfo{title}{Control and communication challenges in networked
    real-time systems}.  \newblock \emph{\bibinfo{journal}{Proc. IEEE}}
  \textbf{\bibinfo{volume}{95}}, \bibinfo{pages}{9--28}
  (\bibinfo{year}{2007}).

\bibitem{olfati-saber07:_consen_cooper_networ_multi_agent_system}
  \bibinfo{author}{Olfati-Saber, R.}, \bibinfo{author}{Fax, J.~A.} \&
  \bibinfo{author}{Murray, R.~M.}  \newblock \bibinfo{title}{Consensus and
    cooperation in networked multi-agent systems}.  \newblock
  \emph{\bibinfo{journal}{Proc. {IEEE}}} \textbf{\bibinfo{volume}{95}},
  \bibinfo{pages}{215--233} (\bibinfo{year}{2007}).

\bibitem{nair03:_expon} \bibinfo{author}{Nair, G.~N.} \&
  \bibinfo{author}{Evans, R.~J.}  \newblock \bibinfo{title}{Exponential
    stabilisability of finite-dimensional linear systems with limited date
    rates}.  \newblock \emph{\bibinfo{journal}{Automatica}}
  \textbf{\bibinfo{volume}{39}}, \bibinfo{pages}{585--593}
  (\bibinfo{year}{2003}).

\bibitem{tatikonda04:_contr} \bibinfo{author}{Tatikonda, S.} \&
  \bibinfo{author}{Mitter, S.~K.}  \newblock \bibinfo{title}{Control under
    communication constraints}.  \newblock \emph{\bibinfo{journal}{IEEE
      Trans. Autom. Control}} \textbf{\bibinfo{volume}{49}},
  \bibinfo{pages}{1549--1561} (\bibinfo{year}{2004}).

\bibitem{nair07:_feedb_contr_under_data_rate_const} \bibinfo{author}{Nair,
    G.~N.}, \bibinfo{author}{Fagnani, F.}, \bibinfo{author}{Zampieri, S.} \&
  \bibinfo{author}{Evans, R.~J.}  \newblock \bibinfo{title}{Feedback control
    under data rate constraints: An overview}.  \newblock
  \emph{\bibinfo{journal}{Proc. IEEE}} \textbf{\bibinfo{volume}{95}},
  \bibinfo{pages}{108--137} (\bibinfo{year}{2007}).

\bibitem{wong97:_system_finit_commun_bandw_const} \bibinfo{author}{Wong,
    W.~S.} \& \bibinfo{author}{Brockett, R.~W.}  \newblock
  \bibinfo{title}{Systems with finite communication bandwidth
    constraints---part {I}: Estimation problems}.  \newblock
  \emph{\bibinfo{journal}{IEEE Trans. Automat. Contr.}}
  \textbf{\bibinfo{volume}{42}}, \bibinfo{pages}{1294--1299}
  (\bibinfo{year}{1997}).

\bibitem{wong97:_system_finit_commun_bandw_const_2} \bibinfo{author}{Wong,
    W.~S.} \& \bibinfo{author}{Brockett, R.~W.}  \newblock
  \bibinfo{title}{Systems with finite communication bandwidth
    constraints---part {II}: Stabilization with limited information feedback}.
  \newblock \emph{\bibinfo{journal}{IEEE Trans. Automat. Contr.}}
  \textbf{\bibinfo{volume}{44}}, \bibinfo{pages}{1049--1053}
  (\bibinfo{year}{1997}).

\bibitem{moreau05:_stabil_multiag_system_with_time} \bibinfo{author}{Moreau,
    L.}  \newblock \bibinfo{title}{Stability of multiagent systems with
    time-dependent communication links}.  \newblock
  \emph{\bibinfo{journal}{IEEE Trans. Automat. Cont.}}
  \textbf{\bibinfo{volume}{50}}, \bibinfo{pages}{169--182}
  (\bibinfo{year}{2005}).

\bibitem{yu10:_secon} \bibinfo{author}{Yu, W.}, \bibinfo{author}{Chen, G.},
  \bibinfo{author}{Cao, M.}  \& \bibinfo{author}{Kurths, J.}  \newblock
  \bibinfo{title}{Second-order consensus for multiagent systems with directed
    topologies and nonlinear dynamics}.  \newblock
  \emph{\bibinfo{journal}{IEEE Trans. on Systems, Man, and Cybernetics-Part
      B}} \textbf{\bibinfo{volume}{40}}, \bibinfo{pages}{881--891}
  (\bibinfo{year}{2010}).

\bibitem{jad} \bibinfo{author}{Jadbabaie, A.}, \bibinfo{author}{Lin, J.} \&
  \bibinfo{author}{Morse, A.~S.}  \newblock \bibinfo{title}{Coordination of
    groups of mobile autonomous agents using nearest neighbor rules}.
  \newblock \emph{\bibinfo{journal}{IEEE Trans. Autom. Contr.}}
  \textbf{\bibinfo{volume}{48}}, \bibinfo{pages}{988--1001}
  (\bibinfo{year}{2003}).

\bibitem{ref:ren} \bibinfo{author}{Ren, W.} \& \bibinfo{author}{Beard, R.}
  \newblock \bibinfo{title}{Consensus seeking in multiagent systems under
    dynamically changing interaction topologies}.  \newblock
  \emph{\bibinfo{journal}{IEEE Trans. Autom. Control}}
  \textbf{\bibinfo{volume}{50}}, \bibinfo{pages}{655--661}
  (\bibinfo{year}{2005}).

\bibitem{scholz98:_compl_block_ttx_resis_na} \bibinfo{author}{Scholz, A.},
  \bibinfo{author}{Kuboyama, N.}, \bibinfo{author}{Hempelmann, G.} \&
  \bibinfo{author}{Vogel, W.}  \newblock \bibinfo{title}{Complex blockade of
    ttx-resistant na$^+$ currents by lidocaine and bupivacaine reduce firing
    frequency in drg neurons}.  \newblock
  \emph{\bibinfo{journal}{J. Neurophysiol.}}  \textbf{\bibinfo{volume}{279}},
  \bibinfo{pages}{1746--1754} (\bibinfo{year}{1998}).

\bibitem{vicsek95:_novel} \bibinfo{author}{Vicsek, T.},
  \bibinfo{author}{Czir{\'o}k, A.}, \bibinfo{author}{Ben-Jacob, E.},
  \bibinfo{author}{Cohen, I.} \& \bibinfo{author}{Shochet, O.}  \newblock
  \bibinfo{title}{Novel type of phase-transition in a system of self-driven
    particles}.  \newblock \emph{\bibinfo{journal}{Phys. Rev. Lett.}}
  \textbf{\bibinfo{volume}{75}}, \bibinfo{pages}{1226--1229}
  (\bibinfo{year}{1995}).

\bibitem{dusenbery92:_sensor_ecolog} \bibinfo{author}{Dusenbery, D.~B.}
  \newblock \emph{\bibinfo{title}{Sensory Ecology: How organisms acquire and
      respond to information}} (\bibinfo{publisher}{W. H. Freeman and Co.},
  \bibinfo{address}{New York}, \bibinfo{year}{1992}).

\bibitem{liao07:_review_of_fish_swimm_mechan} \bibinfo{author}{Liao, J.~C.}
  \newblock \bibinfo{title}{A review of fish swimming mechanics and behaviour
    in altered flows}.  \newblock
  \emph{\bibinfo{journal}{Phil. Trans. R. Soc. B}}
  \textbf{\bibinfo{volume}{362}}, \bibinfo{pages}{1973--1993}
  (\bibinfo{year}{2007}).

\bibitem{komareji13:_resil_contr_dynam_collec_behav}
  \bibinfo{author}{Komareji, M.} \& \bibinfo{author}{Bouffanais, R.}
  \newblock \bibinfo{title}{Resilience and controllability of dynamic
    collective behaviors}.  \newblock \emph{\bibinfo{journal}{PLoS ONE}}
  \textbf{\bibinfo{volume}{8}}, \bibinfo{pages}{e82578}
  (\bibinfo{year}{2013}).

\bibitem{young13:_starl_flock_networ_manag_uncer} \bibinfo{author}{Young,
    G.~F.}, \bibinfo{author}{Scardovi, L.}, \bibinfo{author}{Cavagna, A.},
  \bibinfo{author}{Giardina, I.} \& \bibinfo{author}{Leonard, N.~E.}
  \newblock \bibinfo{title}{Starling flock networks manage uncertainty in
    consensus at low cost}.  \newblock \emph{\bibinfo{journal}{PLoS
      Comput. Biol.}}  \textbf{\bibinfo{volume}{9}}, \bibinfo{pages}{e1002894}
  (\bibinfo{year}{2013}).

\bibitem{fitch13:_infor_centr_optim_leader_selec_noisy_networ}
  \bibinfo{author}{Fitch, K.} \& \bibinfo{author}{Leonard, N.~E.}  \newblock
  \bibinfo{title}{Information centrality and optimal leader selection in noisy
    networks}.  \newblock In \emph{\bibinfo{booktitle}{IEEE 52nd Conference on
      Decision and Control (CDC)}}, \bibinfo{pages}{7510--7515}
  (\bibinfo{publisher}{IEEE}, \bibinfo{year}{2013}).

\bibitem{radakov73:_school_ecolog_fish} \bibinfo{author}{Radakov, D.}
  \newblock \emph{\bibinfo{title}{Schooling in the Ecology of Fish}}
  (\bibinfo{publisher}{John Wiley and Sons}, \bibinfo{address}{New York},
  \bibinfo{year}{1973}).

\bibitem{treherne81:_group} \bibinfo{author}{Treherne, J.} \&
  \bibinfo{author}{Foster, W.}  \newblock \bibinfo{title}{Group transmission
    of predator avoidance behavior in a marine insect: the trafalgar effect}.
  \newblock \emph{\bibinfo{journal}{Anim. Behav.}}
  \textbf{\bibinfo{volume}{29}}, \bibinfo{pages}{911--917}
  (\bibinfo{year}{1981}).

\bibitem{shang14:_consen} \bibinfo{author}{Shang, Y.} \&
  \bibinfo{author}{Bouffanais, R.}  \newblock \bibinfo{title}{Consensus
    reaching in swarms ruled by a hybrid metric-topological distance}.
  \newblock \emph{\bibinfo{journal}{Europ. Phys. J. B}}
  \textbf{\bibinfo{volume}{87}}, \bibinfo{pages}{294} (\bibinfo{year}{2014}).

\bibitem{holme12:_tempor} \bibinfo{author}{Holme, P.} \&
  \bibinfo{author}{Saram\"{a}ki, J.}  \newblock \bibinfo{title}{Temporal
    networks}.  \newblock \emph{\bibinfo{journal}{Phys. Rep.}}
  \textbf{\bibinfo{volume}{519}}, \bibinfo{pages}{97--125}
  (\bibinfo{year}{2012}).

\bibitem{mackay03:_infor_theor_infer_learn_algor} \bibinfo{author}{MacKay, D.}
  \newblock \emph{\bibinfo{title}{Information Theory, Inference, and Learning
      Algorithms}} (\bibinfo{publisher}{Cambridge University Press},
  \bibinfo{year}{2003}).

\bibitem{meyer72:_physic_applied_acous} \bibinfo{author}{Meyer, E.} \&
  \bibinfo{author}{Neumann, E.-G.}  \newblock \emph{\bibinfo{title}{Physical
      and Applied Acoustics}} (\bibinfo{publisher}{Academic Press},
  \bibinfo{address}{New York}, \bibinfo{year}{1972}).

\bibitem{papadimitriou98:_combin_optim} \bibinfo{author}{Papadimitriou, C.} \&
  \bibinfo{author}{Steiglitz, K.}  \newblock
  \emph{\bibinfo{title}{Combinatorial {O}ptimization: {A}lgorithms and
      {C}omplexity}}, chap. \bibinfo{chapter}{6.1 The Max-Flow, Min-Cut
    Theorem}, \bibinfo{pages}{117--120} (\bibinfo{publisher}{Dover
    Publications}, \bibinfo{address}{New York}, \bibinfo{year}{1998}).

\bibitem{wilson62:_chemic_fr} \bibinfo{author}{Wilson, E.~O.}  \newblock
  \bibinfo{title}{Chemical communication among workers of the fire ant
    \emph{Solenopsis saevissima} ({F}r. {S}mith). 1. {T}he organization of
    mass-foraging}.  \newblock \emph{\bibinfo{journal}{Anim. Behav.}}
  \textbf{\bibinfo{volume}{10}}, \bibinfo{pages}{134--147}
  (\bibinfo{year}{1962}).

\bibitem{berger92:_bound} \bibinfo{author}{Berger, M.} \&
  \bibinfo{author}{Wang, Y.}  \newblock \bibinfo{title}{Bounded semigroups of
    matrices}.  \newblock \emph{\bibinfo{journal}{Linear Algebra Appl.}}
  \textbf{\bibinfo{volume}{166}}, \bibinfo{pages}{21--27}
  (\bibinfo{year}{1992}).

\bibitem{horn87:_matrix_analy} \bibinfo{author}{Horn, R.~A.} \&
  \bibinfo{author}{Johnson, C.~R.}  \newblock \emph{\bibinfo{title}{Matrix
      Analysis}} (\bibinfo{publisher}{Cambridge University Press},
  \bibinfo{address}{Cambridge, U.K.}, \bibinfo{year}{1987}).

\bibitem{binder81:_finit_size_scalin_analy_ising} \bibinfo{author}{Binder, K.}
  \newblock \bibinfo{title}{Finite size scaling analysis of ising model block
    distribution functions}.  \newblock \emph{\bibinfo{journal}{Z. Phys. B -
      Condensed Matter}} \textbf{\bibinfo{volume}{43}},
  \bibinfo{pages}{119--140} (\bibinfo{year}{1981}).

\bibitem{gregoire04:_onset_of_collec_and_cohes_motion}
  \bibinfo{author}{Gr\'egoire, G.} \& \bibinfo{author}{Chat\'e, H.}  \newblock
  \bibinfo{title}{Onset of collective and cohesive motion}.  \newblock
  \emph{\bibinfo{journal}{Phys. Rev. Lett.}}  \textbf{\bibinfo{volume}{92}},
  \bibinfo{pages}{025702} (\bibinfo{year}{2004}).

\bibitem{chate08:_collec} \bibinfo{author}{Chat\'e, H.},
  \bibinfo{author}{Ginelli, F.}, \bibinfo{author}{Gr\'egoire, G.} \&
  \bibinfo{author}{Raynaud, F.}  \newblock \bibinfo{title}{Collective motion
    of self-propelled particles interacting without cohesion}.  \newblock
  \emph{\bibinfo{journal}{Phys. Rev. E}} \textbf{\bibinfo{volume}{77}},
  \bibinfo{pages}{046113} (\bibinfo{year}{2008}).

\bibitem{solon14:_from_phase_micro_phase_separ_flock_model}
  \bibinfo{author}{Solon, A.~P.}, \bibinfo{author}{Chat\'e, H.} \&
  \bibinfo{author}{Tailleur, J.}  \newblock \bibinfo{title}{From phase to
    micro-phase separation in flocking models: The essential role of
    non-equilibrium fluctuations} (\bibinfo{year}{2014}).  \newblock
  \bibinfo{note}{ArXiv}, \eprint{1406.6088v2 [cond-mat.stat.mech]}.

\bibitem{viscek2012} \bibinfo{author}{Vicsek, T.} \&
  \bibinfo{author}{Zafeiris, A.}  \newblock \bibinfo{title}{Collective
    motion}.  \newblock \emph{\bibinfo{journal}{Phys. Rep.}}
  \textbf{\bibinfo{volume}{517}}, \bibinfo{pages}{71--140}
  (\bibinfo{year}{2012}).

\bibitem{elia01:_stabil_linear_system_with_limit_infor} \bibinfo{author}{Elia,
    N.} \& \bibinfo{author}{Mitter, S.~K.}  \newblock
  \bibinfo{title}{Stabilization of linear systems with limited information}.
  \newblock \emph{\bibinfo{journal}{IEEE Trans. Automat. Contr.}}
  \textbf{\bibinfo{volume}{46}}, \bibinfo{pages}{1384--1400}
  (\bibinfo{year}{2001}).

\bibitem{glass86:_atlan_scomb} \bibinfo{author}{Glass, C.},
  \bibinfo{author}{Wardle, C.} \& \bibinfo{author}{Mojsiewicz, W.}  \newblock
  \bibinfo{title}{A light intensity threshold for schooling in the {A}tlantic
    mackerel, {\em scomber scombrus}}.  \newblock
  \emph{\bibinfo{journal}{J. Fish Biol.}}  \textbf{\bibinfo{volume}{29}},
  \bibinfo{pages}{71--81} (\bibinfo{year}{1986}).

\bibitem{keenleyside55:_some_aspec_school_behav_fish}
  \bibinfo{author}{Keenleyside, M.}  \newblock \bibinfo{title}{Some aspects of
    the schooling behaviour of fish}.  \newblock
  \emph{\bibinfo{journal}{Behaviour}} \textbf{\bibinfo{volume}{8}},
  \bibinfo{pages}{183--248} (\bibinfo{year}{1955}).

\bibitem{emery73:_prelim_compar_day_night_habit} \bibinfo{author}{Emery, A.}
  \newblock \bibinfo{title}{Preliminary comparisons of day and night habits of
    freshwater fish in ontario lakes}.  \newblock
  \emph{\bibinfo{journal}{Journal of the Fisheries Research Board of Canada}}
  \textbf{\bibinfo{volume}{30}}, \bibinfo{pages}{761--774}
  (\bibinfo{year}{1973}).

\bibitem{Note1} \bibinfo{note}{Equivalent to information update since the
    Nyquist rate $2B=f$ relates bandwidth and frequency of update}.

\bibitem{barberis14:_eviden} \bibinfo{author}{Barberis, L.} \&
  \bibinfo{author}{Albano, E.~V.}  \newblock \bibinfo{title}{Evidence of a
    robust universality class in the critical behavior of self-propelled
    agents: Metric versus topological interactions}.  \newblock
  \emph{\bibinfo{journal}{Phys. Rev. E}} \textbf{\bibinfo{volume}{89}},
  \bibinfo{pages}{012139} (\bibinfo{year}{2014}).

\end{thebibliography}
%

%
\end{document}